\documentstyle[12pt]{article}

\begin{document}

\date{November 19, 1995}

\def\spose#1{\hbox to 0pt{#1\hss}}
\def\ltapprox{\mathrel{\spose{\lower 3pt\hbox{$\mathchar"218$}}
 \raise 2.0pt\hbox{$\mathchar"13C$}}}
\def\gtapprox{\mathrel{\spose{\lower 3pt\hbox{$\mathchar"218$}}
 \raise 2.0pt\hbox{$\mathchar"13E$}}}
\def\inapprox{\mathrel{\spose{\lower 3pt\hbox{$\mathchar"218$}}
 \raise 2.0pt\hbox{$\mathchar"232$}}}

\font\sevenrm  = cmr7 
\def\fancyplus{\hbox{+\kern-6.65pt\lower3.2pt\hbox{\sevenrm --}%
\kern-4pt\raise4.6pt\hbox{\sevenrm --}%
\kern-9pt\raise0.65pt\hbox{$\tiny\vdash$}%
\kern-2pt\raise0.65pt\hbox{$\tiny\dashv$}}}
\def\fancycross{\hbox{$\times$\kern-9.8pt\raise3.6pt\hbox{$\tiny \times$}%
\kern-1.2pt\raise3.6pt\hbox{$\tiny \times$}%
\kern-10.5pt\lower1.0pt\hbox{$\tiny \times$}%
\kern-1.2pt\lower1.0pt\hbox{$\tiny \times$}}}
\def\fancysquare{%
\hbox{$\small\Box$\kern-9.6pt\raise7.1pt\hbox{$\vpt\backslash$}%
\kern+3.9pt\raise7.1pt\hbox{$\vpt /$}%
\kern-11.5pt\lower3.0pt\hbox{$\vpt /$}%
\kern+3.9pt\lower3.0pt\hbox{$\vpt\backslash$}}}
\def\fancydiamond{\hbox{$\diamond$\kern-4.25pt\lower3.8pt\hbox{$\vpt\vert$}%
\kern-2.25pt\raise7pt\hbox{$\vpt\vert$}%
\kern-7.05pt\raise1.57pt\hbox{\vpt --}%
\kern+5.25pt\raise1.57pt\hbox{\vpt --}}}

\title{\vspace{-2cm}
           Reply to Comment on\break
           ``Asymptotic Scaling in the\break
           Two-Dimensional $O(3)$ $\sigma$-Model\break
           at Correlation Length $10^5$''}

\author{
  {\small Sergio Caracciolo}              \\[-0.2cm]
  {\small\it Dipartimento di Fisica}  \\[-0.2cm]
  {\small\it Universit\`a di Lecce and INFN -- Sezione di Lecce}    \\[-0.2cm]
  {\small\it I-73100 Lecce, ITALIA}          \\[-0.2cm]
  {\small Internet: {\tt CARACCIO@LE.INFN.IT}}     \\[-0.2cm]
  \\[-4mm]  \and
  {\small Robert G. Edwards}              \\[-0.2cm]
  {\small\it Supercomputer Computations Research Institute}         \\[-0.2cm]
  {\small\it  Florida State University}   \\[-0.2cm]
  {\small\it Tallahassee, FL 32306 USA}   \\[-0.2cm]
  {\small Internet: {\tt EDWARDS@SCRI.FSU.EDU}} \\[-0.2cm]
  \\[-4mm]  \and
  {\small Andrea Pelissetto}              \\[-0.2cm]
  {\small\it Dipartimento di Fisica and INFN -- Sezione di Pisa}  \\[-0.2cm]
  {\small\it Universit\`a degli Studi di Pisa}  \\[-0.2cm]
  {\small\it I-56100 Pisa, ITALIA}              \\[-0.2cm]
  {\small Internet: {\tt PELISSET@SUNTHPI1.DIFI.UNIPI.IT}}  \\[-0.2cm]
  \\[-4mm]  \and
  {\small Alan D. Sokal}                  \\[-0.2cm]
  {\small\it Department of Physics}       \\[-0.2cm]
  {\small\it New York University}         \\[-0.2cm]
  {\small\it New York, NY 10003 USA}      \\[-0.2cm]
  {\small Internet: {\tt SOKAL@NYU.EDU}}        \\[-0.2cm]
  {\protect\makebox[5in]{\quad}}  
  \\
}

\maketitle
\thispagestyle{empty}   

\vspace{1cm}

\noindent
{\bf PACS number(s):}  11.10.Gh, 11.15.Ha, 12.38.Gc, 05.70.Jk


\clearpage

\newcommand{\be}{\begin{equation}}
\newcommand{\ee}{\end{equation}}
\newcommand{\<}{\langle}
\renewcommand{\>}{\rangle}
\newcommand{\para}{\|}
\renewcommand{\perp}{\bot}

\def\half{ {{1 \over 2 }}}
\def\smfrac#1#2{{\textstyle\frac{#1}{#2}}}
\def\smhalf{ {\smfrac{1}{2}} }
\def\scra{{\cal A}}
\def\scrc{{\cal C}}
\def\scre{{\cal E}}
\def\scrf{{\cal F}}
\def\scrh{{\cal H}}
\def\scrm{{\cal M}}
\newcommand{\scrmvec}{\vec{\cal M}}
\def\scro{{\cal O}}
\def\scrp{{\cal P}}
\def\scrr{{\cal R}}
\def\scrs{{\cal S}}
\def\scrt{{\cal T}}
\def\ttens{{\stackrel{\leftrightarrow}{T}}}
\def\scrttens{{\stackrel{\leftrightarrow}{\cal T}}}
\def\scrv{{\cal V}}
\def\scrw{{\cal W}}
\def\scry{{\cal Y}}
\def\tauss{\tau_{int,\,\scrm^2}}
\def\taux{\tau_{int,\,{\cal M}^2}}
\newcommand{\taum}{\tau_{int,\,\vec{\cal M}}}
\def\taue{\tau_{int,\,{\cal E}}}
\newcommand{\imag}{\mathop{\rm Im}\nolimits}
\newcommand{\real}{\mathop{\rm Re}\nolimits}
\newcommand{\tr}{\mathop{\rm tr}\nolimits}
\newcommand{\sgn}{\mathop{\rm sgn}\nolimits}
\newcommand{\codim}{\mathop{\rm codim}\nolimits}
\def\textprime{{${}^\prime$}}
\newcommand{\longto}{\longrightarrow}
\def\var{ \hbox{var} }
\newcommand{\gtilde}{ {\widetilde{G}} }
\newcommand{\USp}{ \hbox{\it USp} }
\newcommand{\CP}{ \hbox{\it CP\/} }
\newcommand{\QP}{ \hbox{\it QP\/} }
\def\hboxscript#1{ {\hbox{\scriptsize\em #1}} }

\newcommand{\plotdot}{\makebox(0,0){$\bullet$}}
\newcommand{\plotsmalldot}{\makebox(0,0){{\footnotesize $\bullet$}}}

\def\bsigma{\mbox{\protect\boldmath $\sigma$}}
\def\btau{\mbox{\protect\boldmath $\tau$}}
\def\br{{\bf r}}

\newcommand{\reff}[1]{(\ref{#1})}

\font\specialroman=msym10 scaled\magstep1  
\font\sevenspecialroman=msym7              
\def\zed{\hbox{\specialroman Z}}
\def\szed{\hbox{\sevenspecialroman Z}}
\def\R{\hbox{\specialroman R}}
\def\sR{\hbox{\sevenspecialroman R}}
\def\N{\hbox{\specialroman N}}
\def\C{\hbox{\specialroman C}}
\def\Q{\hbox{\specialroman Q}}
\renewcommand{\emptyset}{\hbox{\specialroman ?}}



\newtheorem{theorem}{Theorem}[section]
\newtheorem{corollary}[theorem]{Corollary}
\newtheorem{lemma}[theorem]{Lemma}
\def\proof{\bigskip\par\noindent{\sc Proof.\ }}
\def\qed{\hbox{\hskip 6pt\vrule width6pt height7pt depth1pt \hskip1pt}\bigskip}

%
%
\newenvironment{sarray}{
          \textfont0=\scriptfont0
          \scriptfont0=\scriptscriptfont0
          \textfont1=\scriptfont1
          \scriptfont1=\scriptscriptfont1
          \textfont2=\scriptfont2
          \scriptfont2=\scriptscriptfont2
          \textfont3=\scriptfont3
          \scriptfont3=\scriptscriptfont3
        \renewcommand{\arraystretch}{0.7}
        \begin{array}{l}}{\end{array}}

\newenvironment{scarray}{
          \textfont0=\scriptfont0
          \scriptfont0=\scriptscriptfont0
          \textfont1=\scriptfont1
          \scriptfont1=\scriptscriptfont1
          \textfont2=\scriptfont2
          \scriptfont2=\scriptscriptfont2
          \textfont3=\scriptfont3
          \scriptfont3=\scriptscriptfont3
        \renewcommand{\arraystretch}{0.7}
        \begin{array}{c}}{\end{array}}


As Patrascioiu and Seiler \cite{Pat-Seil_comment} note,
there are two {\em very different}\/
limits that can be taken in a two-dimensional $\sigma$-model:
(a) $\beta\to\infty$ at {\em fixed}\/ $L < \infty$;
or (b) $\beta\to\infty$ and $L\to\infty$ such that the ratio
$x \equiv \xi(\beta,L)/L$ is held fixed.
Limit (b) is the one relevant to finite-size scaling,
while perturbation theory is clearly valid in limit (a).
The deep question is whether the perturbation theory derived
from the study of limit (a) is {\em also}\/ correct in the double limit
obtained by first taking limit (b) and then taking $x\to\infty$.
The conventional wisdom says {\em yes}\/:
indeed, this or a similar interchange of limits
underlies the conventional derivations of asymptotic freedom.
Patrascioiu and Seiler say {\em no}\/:
they suspect that asymptotic freedom is false \cite{Pat-Seil}.
At present, no rigorous proof is available to settle this question
one way or the other.

Our analysis \cite{o3_letter} of our Monte Carlo data
is based on finite-size scaling
\cite{Luscher_91,Kim_93,fss_greedy}, i.e.\ limit (b).
Thus, at each {\em fixed}\/ $x \equiv \xi(\beta,L)/L$, we ask whether
the ratios ${\cal O}(\beta,2L)/{\cal O}(\beta,L)$ have a good limit
as $L\to\infty$, and we attempt to evaluate this limit numerically
in the usual way:  namely, we evaluate the ratios over a wide range of $L$
(from 32 to 256) and we ask whether
these ratios appear to be converging to a limit as $L$ grows.
We find, in fact, that the ratios are {\em constant}\/
within error bars for $L \gtapprox 64$--128 (depending on the value of $x$).
Of course, it is {\em conceivable}\/ that this apparent limiting value
is a deception --- i.e.\ a ``false plateau'' --- and that at much larger values
of $L$ the ratio will change dramatically.  We acknowledge as much
in the penultimate paragraph of our Letter.  This caveat is not special
to our work, but is inherent in {\em any}\/ numerical work which
attempts to evaluate a limit (here $L\to\infty$) by taking the relevant
parameter {\em almost}\/ to the limit (here $L$ large but finite).

In any case, there is no evidence
that this perverse scenario
in fact occurs.
The corrections to scaling in our data are very weak --- less than 2\% even
at $L=32$, and a fraction of a percent or smaller for $L \gtapprox 64$--128 ---
and are perfectly consistent with a behavior of the form
\be
   {{\cal O}(\beta,2L) \over {\cal O}(\beta,L)}
   \;=\;  F_{\cal O}(x) \,+\, {1 \over L^2} G_{\cal O}(x) \,+\, \ldots
\ee
where the correction term $G_{\cal O}$ is negative for
$0.3 \ltapprox x \ltapprox 0.7$ and is perhaps slightly positive
for $x \gtapprox 0.7$.  If all hell breaks loose for larger $L$ ---
as the Patrascioiu-Seiler scenario would require --- we certainly see
no hint of it at $L \le 256$.

Patrascioiu-Seiler also note that our Monte Carlo data at $x \gtapprox 0.7$
agree well with the 2-loop perturbative prediction, shown as a dotted
curve in Figure 2 of \cite{o3_letter}.
But this does not mean that we are {\em assuming}\/
asymptotic scaling (whether explicitly or implicitly).
Quite the contrary:  our data at $x \gtapprox 0.7$ constitute
a (weak) {\em test}\/ of asymptotic scaling.
The same point $(\beta,L)$ may well lie within the range of validity
(to some given accuracy) of two distinct expansions.
The fact that our data points at large $x$ are consistent with
finite-volume perturbation theory [limit (a)]
does not constitute evidence against their {\em also}\/ being consistent with
nonperturbative finite-size scaling [limit (b)].

Of course, since our Monte Carlo data for
$F_{\cal O}(x)$ at $x \gtapprox 0.7$ do in fact agree
closely with the two-loop perturbative formula (to within about 1\%),
and our data for ${\cal O}(\beta,L)$ also agree well with the fixed-$L$
perturbation expansion (to within a few percent),
it is then inevitable that our extrapolated values $\xi_\infty(\beta)$
at the largest values of $\beta$ will be consistent with asymptotic
scaling, in the sense that
$\xi_\infty(\beta)/[e^{2\pi\beta/(N-2)} \beta^{-1/(N-2)}]$
will be roughly constant.
However, it is by no means inevitable that this constant value will agree
with the Hasenfratz-Maggiore-Niedermayer (HMN) prediction to within 4\%.
It seems to us that this apparent coincidence is significant evidence
in favor of the asymptotic-freedom picture.


Finally, Patrascioiu and Seiler \cite{Pat-Seil_superinstanton}
have found an unusual
boundary condition for which the $L\to\infty$ limit of the perturbative
coefficients {\em disagrees}\/ with those obtained from the same limit
in periodic boundary conditions.  Since the two boundary conditions
should agree in the limit $L\to\infty$ at any {\em fixed}\/ $\beta < \infty$,
it follows that {\em for at least one of the two boundary conditions}\/
the $L\to\infty$ limit fails to commute with perturbation expansion
in powers of $1/\beta$.
This is troubling, but it does not tell us {\em which}\/ of the two
boundary conditions is at fault.  It is quite possible that
the two limits {\em do}\/ commute in periodic boundary conditions ---
as the conventional wisdom asserts --- but not in Patrascioiu-Seiler's
unusual boundary condition.  Nevertheless, this example shows that the
justification of the conventional wisdom --- if indeed it is true ---
will be considerably more subtle than was heretofore believed.


%
%
%

\clearpage

\end{document}